\documentclass[
12pt,amssymb]{article}
\usepackage[dvips]{graphicx}
\newcommand{\beq}{\begin{equation}}
\newcommand{\eeq}{\end{equation}}
\newcommand{\bea}{\begin{eqnarray}}
\newcommand{\eea}{\end{eqnarray}}
\def\la{\langle}
\def\ra{\rangle}
\def\al{\alpha}

\def\gam{\gamma}

\def\half{\frac{1}{2}}
\def\nubar{\overline{\nu}}

\def\GMem{G_M}

\def\GENC{G_E^{NC}}
\def\GMNC{G_M^{NC}}
\def\GANC{G_A^{NC}}
\def\GACC{G_A}

\def\GAs{G_A^s}

\def\GMs{G_M^s}

\def\Acal{\mathcal{A}}

\begin{document}
  \mbox{} \vspace*{2.5\fill} {\Large\bf
\begin{center}
%
   Looking for strangeness with neutrino-nucleon scattering
%
\end{center}
}
\begin{center}
  {\large W.M. Alberico$^{1}$, S.M. Bilenky$^{2}$, 
    and C. Maieron$^{3}$ }
\end{center}

\begin{small}
\begin{center}
  $^{1}${\sl Dipartimento di Fisica Teorica, Universit\`a di Torino\\
    and
    INFN, Sezione di Torino,
    Via P. Giuria 1, 10125 Torino, Italy
  }\\[2mm]
  $^{2}${\sl 
Joint Institute for Nuclear Research, Dubna, Russia
  }\\[2mm]
  $^{3}${\sl
INFN, Sezione di Catania, Via S. Sofia 64, I-95123 Catania, Italy }
\end{center}
\end{small}

\kern 1. cm \hrule \kern 3mm

\begin{small}\noindent

  {\bf Abstract} \vspace{3mm}
The possibility to determine the
axial strange form factor of the nucleon from elastic neutrino-nucleon
scattering experiments is studied. The existing experimental 
information is shortly mentioned and several  observables which 
could be measured in the near future at new neutrino facilities are 
discussed.
\vspace{1cm}
\hrule

\kern 2mm

\end{small}


\vskip 1 cm

\section{Introduction}
The measurement of the cross-sections for neutral current (NC) neutrino 
(antineutrino) nucleon elastic scattering
\begin{equation}
\nu_{\mu} (\nubar_{\mu}) + N \longrightarrow
\nu_{\mu} (\nubar_{\mu}) + N
\label{nuelas}
\end{equation}
has been indicated as a key tool for the determination of the so-called 
strange form factor of the nucleon, namely the matrix element of the 
(isoscalar) strange axial current:
\bea
\la p'|\bar{s}\gam^\al\gam^5 s|p\ra = 
\bar{u}(p')\gam^\al\gam^5 u(p)G_A^s(Q^2)
\nonumber
\eea
Here $s,\bar{s}$ is the strange quark field operator, 
$|p\ra$ ($|p'\ra$) is the state vector of a nucleon with 
momentum $p$ ($p'$). 
The NC  which intervenes in the process (\ref{nuelas}) is:
\beq
J_\al^Z=V_\al^3+A_\al^3-2\sin^2\theta_W J^{em}_\al-
\frac{1}{2}V^s_\al-\frac{1}{2}A^s_\al\,.
\label{Jneutral}
\eeq
It contains the customary vector and axial isovector components, which
are the third components of isovectors, the electromagnetic current and
the strange (isoscalar) axial and vector currents (though heavier 
quarks could come into play as well).

In order to disentangle the tiny effect (the present estimates are of the 
order of a few $\%$) of the strange form factors, various observables
have been suggested, all of them being {\em ratios} of cross sections: 
these quantities have the advantage of minimizing the uncertainties connected
with, e.g., the determination of the $\nu$-flux and/or the influence of the
nuclear medium, when $\nu~(\bar{\nu})$ are scattered off nuclei\cite{ABM}.

Let us thus consider the following quantities:
\begin{enumerate}
\item NC over CC ratio:
\begin{equation}
R_{NC/CC}(Q^2)= 
\frac{\displaystyle{\left({d\sigma}/{dQ^2}\right)^{NC}_{\nu}}}
{\displaystyle{\left({d\sigma}/{dQ^2}\right)^{CC}_{\nu}}}
\label{NCratio}
\end{equation}
\item
Neutrino-antineutrino asymmetry:
\begin{equation}
\Acal(Q^2)= \frac{\displaystyle{\left(\frac{d\sigma}{dQ^2}\right)^{NC}_{\nu} -
\left(\frac{d\sigma}{dQ^2}\right)^{NC}_{\nubar}}}
{\displaystyle{\left(\frac{d\sigma}{dQ^2}\right)^{CC}_{\nu} -
\left(\frac{d\sigma}{dQ^2}\right)^{CC}_{\nubar}}}\,
\label{nuasymm}
\end{equation}
%
%
%
\end{enumerate}
The cross sections in the denominators refer to the charged current (CC) 
processes:
\begin{equation}
\begin{array}{l}
\nu_{\mu}+n\longrightarrow \mu^-+p\,,\\
\nubar_{\mu}+p\longrightarrow \mu^+ +n
\end{array}
\label{nuquasiel}
\end{equation}
and can be determined with higher accuracy, since all 
particles in the final state can, in principle, be detected, while the final 
$\nu~(\bar{\nu})$ in the NC process (\ref{nuelas}) is not observed.

The elastic NC $\nu~(\bar{\nu})$-nucleon scattering can be written in the
form:
\bea
&&\left(\frac{d\sigma}{dQ^2}\right)^{NC}_{\nu(\nubar)} = 
\frac{G_F^2}{2\pi}\left[\half y^2(\GMNC)^2 
  +\left(1-y-\frac{M}{2E}y\right)
\frac{\displaystyle{(\GENC)^2+\frac{E}{2M}y(\GMNC)^2}}
{\displaystyle{1+\frac{E}{2M}y}}+\right.
\nonumber\\
&&\,\,\,\quad
\left. +\left(\half y^2+1-y+\frac{M}{2E}y\right)(\GANC)^2
\pm 2y\left(1-\half y\right)\GMNC\GANC\right]\,.
\label{crosselas}
\eea
In the above formula
$y={p\cdot q}/{p\cdot k} ={Q^2}/{2p\cdot k}$, 
 $E$ is the $\nu (\bar\nu)$ energy in the laboratory system, 
$p~(k)$ the initial nucleon (neutrino) 
four-momentum, $M$ is the nucleon mass and $Q^2=-q^2$ the square of the
four-momentum transfer. 
Moreover $G_E^{NC}$,  $G_M^{NC}$,  $G_A^{NC}$ are the electric, magnetic 
and axial weak NC form factors of the nucleon, all of them containing 
an isoscalar strange component. In particular
\beq
G_A^{NC;p(n)}(Q^2)=\pm\half \GACC(Q^2) -\half {G_A^s(Q^2)}\,,
\label{Ganc}
\eeq
where the strange axial form factor can be parameterized with the usual
dipole form $G_A^s(Q^2)={g_A^s}/{(1+Q^2/M_A^2})^2$, with $g_A^s=G_A^s(0)$.

From Equation (\ref{crosselas}) it is clear that to the NC scattering 
process several unknown quantities contribute: in particular there are  
three (electric, magnetic and axial) strange form factors, two of which 
($G_M^s$ and $G_A^s$) can produce contributions of similar size. Moreover
it has been pointed out\cite{BNL} that the present uncertainty on the 
axial cutoff mass, $M_A=1.032\pm 0.036$~GeV, allows one to obtain equally 
good fits to the elastic $\nu-N$ scattering cross sections with values of
$|g_A^s|$ ranging from 0 to 0.25. 

\section{The $\nu-\bar\nu$ Asymmetry}

From the explicit evaluation of the NC and CC $\nu~(\bar\nu)$-nucleon 
cross sections, one can express the neutrino-antineutrino asymmetry as 
follows\cite{ABGM}:
\beq
\Acal_{p(n)}=\frac{1}{4}\left(\pm 1-\frac{\GAs}{\GACC}\right)
\left(\pm 1-2\sin^2\theta_W\frac{{\GMem}^{p(n)}}{\GMem^3} 
-\half\frac{\GMs}{\GMem^3}\right)\,,
\label{asymm1}
\eeq
where the $+$ ($-$) sign refer to proton (neutron) respectively.
Taking into account only terms which linearly  depend on the strange
form factors: 
\begin{equation}
\Acal_{p(n)}=\Acal_{p(n)}^0 \mp\frac{1}{8}\frac{{\GMs}}{\GMem^3} 
\mp \frac{{\GAs}}{\GACC}{\Acal_{p(n)}^0}
\label{asymm2}
\end{equation}
we find out that any deviation with respect to the (known) term,
\begin{equation}
\Acal_{p(n)}^0= \frac{1}{4}\left(1\mp 
2\sin^2\theta_W\frac{{\GMem}^{p(n)}}{\GMem^3} \right),
\label{asymm0}
\end{equation}
must be ascribed to a non-vanishing contribution of $G_M^s$ and/or
$G_A^s$.

Measurements of the asymmetry (\ref{asymm1}) is a quite demanding task 
from the experimental point of view, since it requires both $\nu$ and 
$\bar\nu$ beams of comparable intensity. An indirect ``experimental'' 
value of this asymmetry with flux-averaged cross sections was 
extracted~\cite{ABBCCM} from the data of the BNL-734 experiment~\cite{BNL}.
We refer the reader to ref.~\cite{ABBCCM} for the details of the analysis.
The main conclusion, however, was that the present experimental uncertainty
is compatible with any value of $g_A^s$, in the range $0\ge g_A^s\ge -0.12$. 

From Equation~(\ref{asymm2}), the interference 
between the magnetic ($G_M^s$) and axial ($G_A^s$) strange form factors
 is evident: should they have the same sign, then their effects on the
asymmetry get enhanced. The opposite is true, however, if they have opposite
sign.

\section{Future perspectives}

\begin{figure}[t]
\vspace{-1.5cm}
\includegraphics[width=0.8\textwidth]{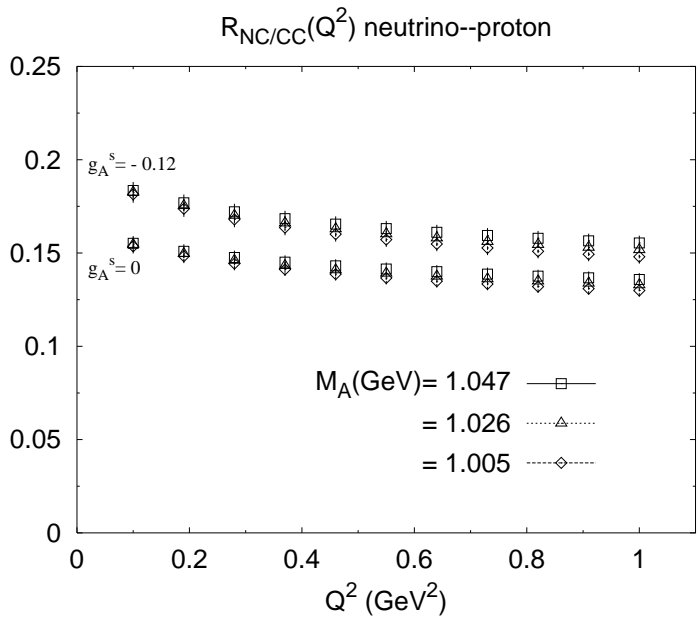}
\vspace{-0.5cm}
\caption{
Plot of the ratio $R_{NC/CC}(Q^2)$, obtained with 
the neutrino cross sections averaged over the $\nu$ spectrum,
for different choices of $M_A$ and $g_A^s$.
}
\label{fig1}
\end{figure}

We consider here the ratio of NC to CC elastic $\nu-p$ scattering cross
sections: the information on the strange form factors one can extract 
from this quantity is not free from ambiguities, however it deserves to be
carefully considered. It was recently proposed~\cite{Rex} to use the high
intensity Booster neutrino beam at Fermilab, to measure $\nu$-nucleon CC 
quasi-elastic and NC elastic scattering, with neutrino energies in the
$0.5\div 1.0$~GeV range. This kinematical conditions appear to be quite
interesting to analyze the ratio $R_{NC/CC}(Q^2)$, Equation (\ref{NCratio}).
From a throughout analysis we have performed on this quantity, 
we can summarize the following outcomes:

\begin{enumerate}
\item
It is sensitive to $g_A^s$, but not much affected by the cutoff mass of
the axial form factors, assumed in the above quoted dipole form.
\item
The interference between axial and vector strange form factors (in particular
the magnetic strange one) can hinder the effect of $g_A^s$ alone. However 
$G_M^s$ is under investigation also with polarized electron-proton scattering
experiments~\cite{Sample} and one can hope to have complementary information
from this source.
\item 
The sensitivity to the flux is negligible, because it is largely eliminated 
in the ratio of cross sections.
\item
The same argument applies to nuclear medium effects: indeed a large fraction
of processes would occur on $^{12}$C, where nucleons are bound and subject
to final state interactions. These  sizeably reduce the single cross 
sections, but their net effect on the ratio  can be safely neglected.
\end{enumerate}

To illustrate some of the above points, we show in fig.~\ref{fig1}
 the NC/CC ratio, for different choices
of the axial cutoff mass $M_A$ and of the strange axial 
constant $g_A^s$, as indicated. We have assumed that
this ratio could be measured with a 5\% accuracy, represented by the
small ``error band'' plotted for each calculated point.
We can see that, for the moderate $Q^2$ values represented here, 
the sensitivity of this ratio to $G_A^s$ is large enough
to allow a precise determination of it. 

We conclude by observing that $\bar\nu$ scattering, if feasible, would
offer relevant and complementary information on the strange form factors
of the nucleon, and, eventually, would allow the determination of the 
neutrino asymmetry.

\end{document}